\documentstyle[12pt,aasms4,psfig]{article}

\def\lsim{\lower.5ex\hbox{$\; \buildrel < \over \sim \;$}}
\def\gsim{\lower.5ex\hbox{$\; \buildrel > \over \sim \;$}}

\def\t{\ifmmode {\tau} \else $\tau$ \fi}

\def\cm{\ifmmode {\rm cm}^{-1} \else cm$^{-1}$ \fi}
\def\s{\ifmmode {\rm s}^{-1} \else s$^{-1}$ \fi}
\def\cc{\ifmmode {\rm cm}^{-3} \else cm$^{-3}$ \fi}
\def\cs{\ifmmode {\rm cm}^{-2} \else cm$^{-2}$ \fi}
\def\g{\ifmmode \gamma \else $\gamma$\fi}
\def\l{\ifmmode \lambda \else $\lambda$\fi}
\def\ls{$\lambda$~}
\def\ts{$\tau$~}
\def\t{\ifmmode \tau \else $\tau$\fi}
\def\G{\ifmmode \Gamma \else $\Gamma$\fi}
\def\Gt{\ifmmode \tilde{\Gamma} \else $\tilde{\Gamma}$\fi}

\def\kms{\ifmmode {\rm km\ s}^{-1} \else km s$^{-1}$\fi}

\def\sc{Schwarzschild~}

\begin{document}

\title{Modeling the X-ray -- UV Correlations in NGC 7469 }

\author{Andrew J. Berkley\altaffilmark{1} Demosthenes 
Kazanas\altaffilmark{2}\& Jonathan Ozik\altaffilmark{1}}

\altaffiltext{1}{Department of Physics, University of Maryland,
College Park, MD 20743}
\altaffiltext{2}{LHEA, NASA/GSFC Code 661, Greenbelt, MD 20771}

\font\rom=cmr10
\centerline{\rom submitted to Astrophys. J.}

\baselineskip=15pt

\begin{abstract}
\noindent We model the correlated X-ray -- UV observations of NGC 7469,
for which well sampled data in both these bands have been obtained 
recently in a multiwavelength monitoring campaign. To this end we derive 
the transfer function in wavelength \ls and time lag \t, for reprocessing 
hard (X-ray) photons from a point source to softer ones (UV-optical) by 
an infinite plane (representing a cool, thin accretion disk) located 
at a given distance below the X-ray source, under the assumption 
that the X-ray flux is absorbed and emitted locally by the disk as 
a black body of temperature appropriate to the incident flux. Using the 
observed X-ray light curve as input we have computed the expected continuum 
UV emission  as a function of time at several wavelengths (\l \l 1315 \AA,
\l \l 6962 \AA, \l \l 15000 \AA, \l \l 30000 \AA) assuming that the 
X-ray source is located one \sc radius above the disk plane, with the 
mass of the black hole $M$ and the latitude angle $\theta$ of the observer
relative to the disk plane as  free parameters. We have searched the 
parameter space of black hole masses and observer azimuthal angles but 
we were unable to reproduce UV light curves which would resemble, even 
remotely, those observed. We also explored whether particular combinations
of the values of these parameters could lead to light curves whose statistical properties (i.e. the autocorrelation and cross correlation functions) would 
match those corresponding to the observed UV light curve at \l \l 1315 \AA. 
Even though we considered black hole masses as large as $10^9$ M$_{\odot}$ 
no such match was possible. 
Our results indicate that some of the fundamental assumptions of this 
model will have to be modified to obtain even approximate agreement
between the observed and model X-ray -- UV light curves.

\end{abstract}


\section{Introduction}


\baselineskip 15pt

The study of the physics of Active Galactic Nuclei (AGN) involves length 
scales much too small to be resolved by current technology or 
technology of the foreseeable future. As a result, this study is conducted 
mainly through the theoretical interpretation of the
spectral and temporal properties of these systems, much in the way
that the study of spectroscopic binaries has been used to deduce the 
properties of the binary system members and the elements of their
orbit. Thus, studies of the spectral energy distribution in AGN have 
revealed the ubiquitous presence of a broad quasithermal component 
in the optical -- UV part of the spectrum, the so called Big Blue Bump (BBB), 
as well as soft and hard ($\sim 10$ keV) X-ray emission, which in cases of 
sufficiently bright objects, was found to extend to several hundred keV. 
At lower frequencies, AGN were found to emit roughly the same amount of 
luminosity in the IR and the far-IR part of the spectrum as in the
higher energy bands. This rough equipartition of the AGN 
luminosity from the far-IR to the X-ray part of the electromagnetic 
spectrum is an interesting fact for which no apparent, compelling 
explanation is presently at hand. 

It was proposed long ago (Shields 1978) that a feature such as the BBB
would signify the presence of a
geometrically thin, optically thick accretion disk. 
It is generally thought that these disks radiate away the locally dissipated 
accretion energy in black body form at the temperature required to  
deliver the necessary radiant flux. For a quasar of luminosity $L \simeq
10^{46} \, L_{46}$ erg \s, associated with a black hole of mass $M = 10^8
M_8 \; M_{\odot}$ the corresponding disk temperature is $T \simeq 10^5 \, 
L_{46}^{1/4} \, M_8^{1/2}$ in reasonable agreement with the observed
excess flux at the  relevant wavelengths. The successful, detailed 
fits of the BBB feature as described above (Malkan \& Sargent 1982; 
Malkan 1983; Sun \& Malkan 1989; Laor \& Netzer 1989), has convincingly 
established the identification of this spectral component with this 
specific structure of accretion flow onto the black hole. Even though 
certain of its properties do not completely conform with this notion, 
(e.g. the absence of prominent Ly $\alpha$ edges, and the magnitude 
of polarization with wavelength (see Koratkar \& Blaes 1999), it is 
generally considered that this spectral component does indicate the
presence of a geometrically thin, optically thick disk in the vicinity
of the AGN black hole. 

The X-ray emission must originate in a tenuous, hot ($kT \sim 10^8 - 
10^9$ K) plasma, and its spectrum has been modeled successfully 
by Comptonization of soft photons by the hot electrons. The source 
of the soft photons is usually not specified but it is (naturally)
considered to be the UV emission of the BBB, while the hot electrons 
are considered to be located in a corona overlying the BBB thin disk. 
This corona is thought to be confined and powered either by magnetic 
loops threading this disk, much the way it is the case with the solar 
corona (Galeev, Rossner \& Vaiana 1979), or to be part of an Advection  
Dominated Flow (Narayan \& Yi 1994) which are hot on their own right. 
The precise arrangement of 
these two components is not well defined; however, the form of the
AGN spectral luminosity distribution  between the optical--UV (BBB) 
and the X-ray bands suggests that 
only a small ($\sim 20\%$) fraction of the soft BBB photons traverse 
the volume occupied by the overlying hot electrons. This fact 
then suggests that either: (a) the hot plasma is confined in a 
small region with extent of only a few \sc radii around the black 
hole (while the BBB emission comes from a much larger region) or 
(b) the X-ray emitting plasma consists of small patches which cover
only partially the thin disk (the source of BBB soft photons).

The sites of the IR and far-IR emission are thought to be at much 
larger distances from the compact object than that of the optical - 
UV emission, a conclusion reached on the basis of the much longer
variability time scales in these components (Edelson \& Malkan 1987) 
compared to those associated with the UV and the X-ray emission in AGN. 
In the context of unified AGN models, it is considered that the IR
and far-IR emission results from reprocessing of the ionizing 
continuum by a molecular torus, whose inclination to the observer's line 
of sight is believed responsible for the apparent dichotomy of Seyfert 
galaxies in types I and II. 

While the above inferences and associations of the various spectral 
AGN components with specific spacial structures of rather well defined 
properties and location appear reasonable and make sense in the broader 
context of AGN physics, one has to bear in mind that they have been 
deduced mainly on the basis of spectral fits and radiative transfer.  
In this respect, it is well known (but some times not appreciated) 
that radiative transfer provides, generally, information only about 
column densities and optical depths (see e.g. Kazanas, Hua \& Titarchuk
1997). However, in order to probe the dynamics and geometry of AGN one 
needs information about physical lengths and densities; these latter 
quantities have to be obtained by independent means, usually from 
time variability. For example, the goal of AGN monitoring (Netzer \&
Peterson 1987) has been precisely this, namely the determination 
of the physical size of the AGN Broad Line Region (BLR) using the
reverberation mapping technique. The results of this effort have 
shown that the size of specific AGN components (the BLR region in 
this case) can in fact be very different from prior estimates based 
on spectral considerations alone (the size of BLR was found to 
be off by a factor of $\sim 10$ and the cloud density by a factor 
of $\sim 100$; for a review see Netzer \& Peterson 1997). 

An additional product of the extensive AGN multiwavelength monitoring 
efforts - which were aimed primarily in the determination of the 
BLR size from the continuum- Broad Emission Line correlations - 
has been the measurement of lags in the cross  correlation functions
between the optical and the UV continua. These are important because 
both these bands are part of the BBB and are hence thought 
to be produced by the putative thin accretion disk responsible for 
the emission of this spectral component. These studies indicated 
that the optical 
and the UV continua vary with much higher synchrony than expected 
on the basis of simple accretion disk models for the BBB emission. 
In the simplest models the variations in these two components should 
be propagating from the low to the high frequencies (the sense of 
mass inflow) and should be of order of the viscous time scales at 
the appropriate radii. Much shorter time scales are those associated 
with the  propagation of sound waves traveling on the surface 
of the disk (in this case the UV variation could preceed that of the 
optical). Assuming a gas temperature $T \simeq 10^4 - 10^5$ K, typical 
of the values needed to account for the BBB spectral characteristics and
a size $R \simeq 10^{14}$ cm, the sound crossing time scales are of 
order $\tau \sim R / c_s \sim 10^{7.5} - 10^8$ s, much 
longer than the $\simeq 1 ~{\rm day} \simeq 10^5$ s measured 
(Collier et al. 1998) or upper limits (Krolik et al. 1991) to the lags 
between the optical and UV wavelengths. 

For the above reasons it was conjectured (Krolik et al. 1991) that 
the correlated optical -- UV variability maybe caused by the 
reprocessing of X-rays, emitted by the hot corona overlying the thin
disk, since this process yields signals which propagate much faster
(at the speed of light) and might thus account for both the spectral 
and the temporal properties of these systems. 
However, the absence (until recently) of well coordinated, simultaneous 
observations in the X-ray and the UV -- optical bands left this 
conjecture supported only by circumstantial evidence. In fact, timing 
studies and modeling limited to the correlations between UV and 
optical bands gave results consistent with such a picture: Rokaki 
\& Magnan (1992) analyzed the results of the NGC 5548 monitoring 
campaign assuming that the observed variability is due to the 
reprocessing of an unseen harder spectral component (EUV - X-rays) 
by an optically thick geometrically 
thin disk. They found that the correlations between the continuum light 
curves at \l \l 1360, 1840, 2670 and 4870 \AA~ are consistent with their
assumption, provided that the X-ray source had a variability they 
themselves prescribed and it were located 15 $R_S$ above the 
plane of an  accretion disk around a black hole of mass $6 \times 10^7$ 
M$_{\odot}$. Nonetheless, the absence of simultaneous X-ray observations, 
left this entire effort at the level of ``reasonable conjecture".

The launch of RXTE and the simultaneous presence of IUE in orbit 
made such  coordinated observations possible: The active nucleus
NGC 7469 was observed simultaneously both in the  optical (Collier 
et al. 1998), in the UV (Wanders et al. 1997), and in the X-ray 
(Nandra et al. 1998) bands over an interval of roughly thirty days 
with a sampling rate no smaller than once every other orbit (in the 
UV and X-rays). 
The results of this campaign were rather startling and to some extent 
disappointing: While both the X-ray and the UV band exhibited variability 
of similar amplitudes ($\sim 50\%$), there were no apparent, easily 
understood correlations between the variability of these two bands, at least 
none that could be attributed reasonably to reprocessing of the X-rays 
as the cause of the observed optical -- UV variability.
In the UV, Wanders et al. (1997), measured lags of 0.23, 0.32 and 0.28
days between the variations at \l1315 \AA~ and \l\l 1485, 1740 and 1825
\AA~ respectively. With an error of 0.07 days determined through 
Monte Carlo simulations they were unable to decide conclucively
whether these results represented variability in accordance with 
accretion disk models or variability due to contamination by a 
very broad delayed emission
feature which becomes stronger toward the red part of the spectrum.
The optical observations of Collier et al. (1998), in conjuction with 
those of the UV provided a much larger dynamic range in wavelength
which allowed the determination of lags between the UV (\l1315 \AA) 
and the optical (\l\l4815, 6962 \AA) with greater confidence.
The $\sim 2$ day {\it relative} lags between \l1315 and \l6962 \AA~ 
measured can in fact be intepreted as due to reprocessing by an 
accretion disk, as generally considered; however, it is not apparent 
which part of the spectrum drives the observed variability, while 
there appears to be, in addition, an inconsistency by more than a 
factor of 10 between the observed an inferred luminosity of this 
specific model.

Motivated by these observations we have decided to take a closer look
at this particular question through the detailed modeling of the
specific situation thought to take place in the innermost regions
of AGN. Our approach is straightforward and similar in spirit to 
the analysis of Rokaki \& Magnan (1992) : In \S 2 we compute the 
response function of reprocessing  X-rays, from a point source above 
an infinite plane,  as a function of the wavelength of the reprocessed
radiation \l, the lag time \ts and the latitude angle of the observer 
with respect to the disk plane $\theta$. In \S 3 we fold this 
response function with the observed X-ray light curve to produce 
model UV, optical and IR  light curves for different values of the black 
hole mass and inclination angle. We then compute the autocorrelation
and cross correlation functions of the model UV, optical and infrared 
light curves with that of the input X-ray and compare them to those 
observed. We search the black hole mass $M$ - latitude angle $\theta$ 
parameter space in search of combinations which would 
result in autocorrelation and cross correlation functions similar 
to those observed. Finally, in \S 4 the results are summarized and 
conclusions are drawn.

\section{The Response Function}

In order to make the problem of X-ray reprocessing from a hot corona
into UV-optical radiation by an underlying cool, geometrically thin,
optically thick accretion disk as tractable as possible we have made the 
following idealizations: We have assumed the source of X-rays to be
point-like and located at a height $R_X$ above an infinite plane 
representing the accretion disk producing the UV - optical emission 
associated with the BBB. We believe that the above assumptions 
approximate adequately the situation under consideration in that 
the X-ray source does not cover completely the source of 
UV photons, as required by the spectral fits and discussed in 
the introduction. This assumption is furthermore  confirmed 
{\it a posteriori} by the more rapid 
variability of the X-ray relative to the UV emission.

\begin{figure*} [H]
\centerline{\psfig{file=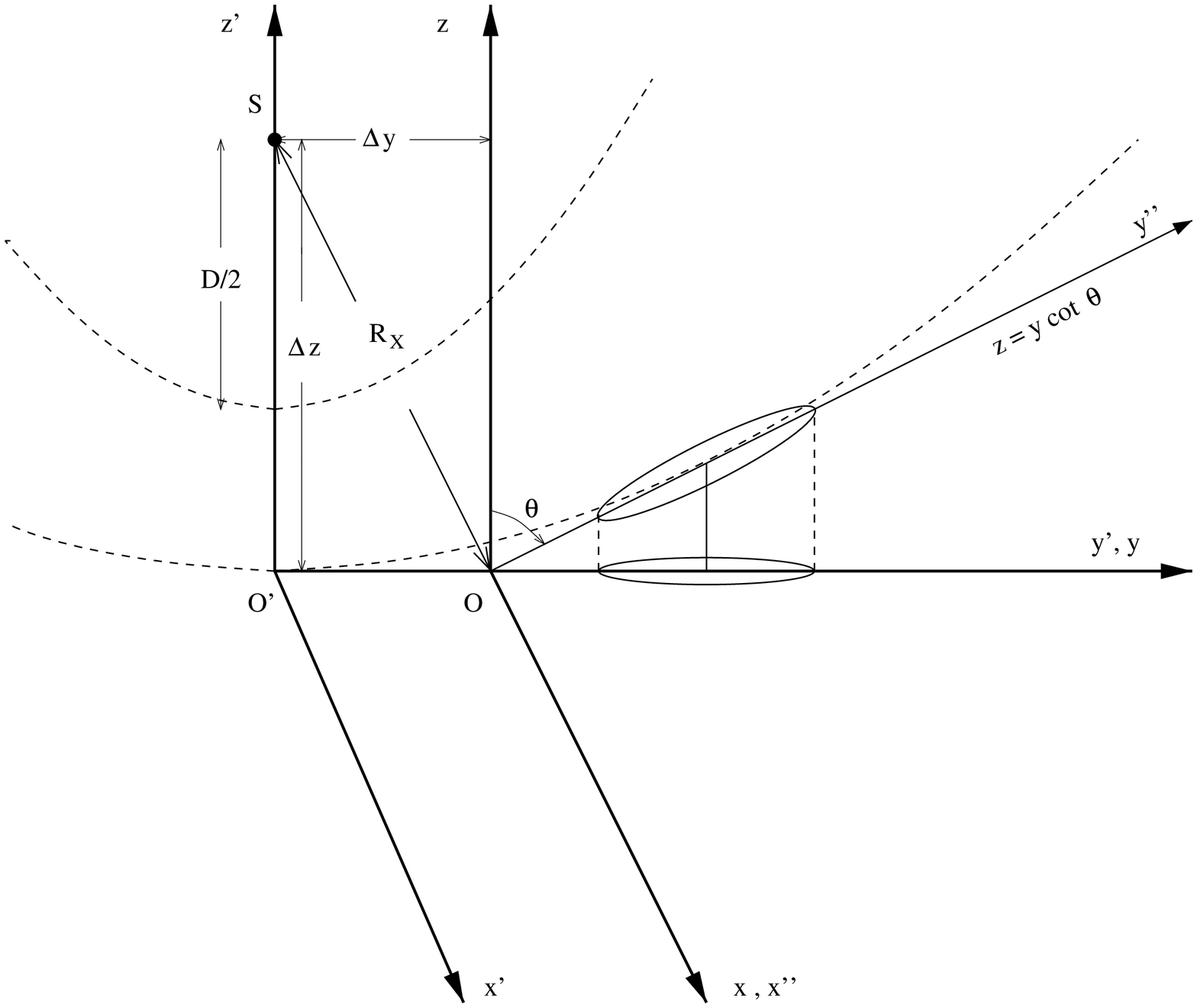,width=.50\textwidth}}
\caption{The geometric arrangement of the source $S$ the accretion 
disk ($x^{\prime \prime},y^{\prime \prime}$ plane), the compact 
object $O$  and the observer (see text). $R_X$ is the height of the
X-ray source above the disk and $\theta$ the latitude of the 
observer with respect to it.} 
\label{geometry}
\end{figure*}

The geometric construction associated with the arrangement of the 
X-ray source and the accretion disk described above is given 
in Figure \ref{geometry}. We choose Cartesian coordinates 
$x^{\prime \prime}, ~y^{\prime \prime}, ~z^{\prime \prime}$
with the $x^{\prime \prime}, ~y^{\prime \prime}$ axes on the plane of 
the disk, their origin  $O$ on the compact object (black hole), 
and the X-ray source $S$ at a distance $R_X$ above the disk plane
in the $z^{\prime \prime}$ direction. The observer's line of 
sight lies on the $y^{\prime \prime}, ~z^{\prime \prime}$ plane, 
along the $z$-direction, as shown  in the figure,  making an angle 
$\theta$ with the plane of the disk. 

The response function $\Psi(\l,\t)$ of the situation depicted in Figure 
\ref{geometry} is obtained from the following considerations: The 
loci of a given constant temperature on the surface of the disk, 
due to the reprocessing of X-rays from the source $S$, are circles 
centered around the foot $O$ of the vertical from the X-ray source 
to the disk plane; on the other hand, the loci 
of constant delay \ts between the X-ray source and the observer are
paraboloids of revolution about the observer - X-ray source 
axis (the $z^{\prime}$ axis) with the X-ray source as their focal 
point.  The intersection of these paraboloids with the plane of 
the disk are generally ellipses;  
the response function $\Psi(\l,\t)$ is precisely the (thermal)
emission by the intersection 
of the ellipses of constant delay with the circles of 
constant temperature on the disk. Because it is assumed that 
the reprocessed radiation is emitted in black body form at 
a temperature determined by the local X-ray flux, the determination
of the response function  $\Psi(\l,\t)$ reduces to computing the 
area of intersection of the constant delay ellipses with the circles 
of constant temperature.

To determine this quantity, we choose Cartesian coordinates 
$x^{\prime}, ~y^{\prime},  ~z^{\prime}$ whose origin is located
on the extremum of the paraboloids of revolution (which have 
the source $S$ as their focal point) of a given constant delay 
$D = c\t$, with the $z^{\prime}$-axis pointing to the 
observer, also located on the $y^{\prime},~z^{\prime}$ plane. 
In these coordinates one can easily verify that the surfaces of 
constant delay $D = c\t$ are paraboloids of revolution of the 
form
\begin{equation}
z^{\prime} = \frac{1}{2D}(x^{\prime \, 2} + y^{\prime \, 2})
\label{parabol}
\end{equation}

\noindent with  the distance between 
the origin of the coordinates (at the extremum of the paraboloids) and 
the X-ray source, located at their focal point, being $D/2$.

We now consider a Cartesian coordinate system $x, ~y, ~z$ with its
origin $O$  located on the compact source and with the $y$ and $z$ 
axes parallel to $y^{\prime}, ~z^{\prime}$ as shown in Figure 
\ref{geometry}.  Let $\Delta y$ and $\Delta z$
be the coordinates of the X-ray source position in this system. 
Then, the relation between the coordinates $x^{\prime }, y^{\prime},
~z^{\prime}$ and  $x, ~y, ~z$ can be easily obatained from the geometric 
construction of Figure \ref{geometry}. These are
\begin{equation}
z^{\prime} = z - (\Delta z - D/2), ~~~~~~~~~y^{\prime} = y +  \Delta y, ~~~
~~~~~~ x^{\prime} = x
\end{equation}
\noindent In terms of these coordinates the equation of paraboloids of 
constant delay (Eq. \ref{parabol}) reads
\begin{equation}
z^{\prime} = z - (\Delta z - D/2) = \frac{1}{2D}[ x^2 + (y +  \Delta y)^2]
\label{para}
\end{equation}
One can now express the coordinates of the source's position 
$\Delta y$ and $\Delta z$ in terms of the height of the source above
the  disk $R_X$ and  the observer's latitude $\theta$, which read:
\begin{equation}
\Delta y = R_X \, cos \,\theta ~~~~~~,~~~~~~ \Delta z = R_X \, sin\, \theta 
\end{equation}
The intersection of the paraboloids of constant delay with the disk 
plane (i.e. the isodelay curves on the disk plane) are obtained 
by finding the intersection of the paraboloid (Eq. \ref{para}) with 
the disk plane $ z =  y \, cot \, \theta$. 
This leads to 
\begin{equation}
y \, cot \, \theta = R_X \, sin\, \theta - \frac{D}{2} + 
\frac{1}{2D}[ x^2 + (y + R_X \, cos \,\theta)^2] ~,
\end{equation}
which after some rearrangement reads
\begin{equation}
\left[y -  (D \, cot \, \theta - R_X \, cos \, \theta)\right]^2  +  x^2 =
\frac{D}{sin \, \theta} \left( \frac{D}{sin \, \theta} - 2 R_X \right)
\label{circle}
\end{equation}
This is the equation of a circle on the plane $x~,y$ (the plane 
perpendicular to the observer's line of sight 
to the X-ray source) of radius square
$D^2/sin^2\theta - 2 \, D \, R_X /sin \, \theta$, which, along with the 
equation for the accretion disk plane $ z =  y \, cot \, \theta$, give
the parametric equations of the sought curve. Clearly the radius
of the circle is non-zero only for sufficiently long lags, i.e. for
$D > 2\, R_X \, sin \, \theta$, that is for the time it takes the 
corresponding isodelay surface to ``cut" the surface of the accretion
disk. For $D = 2\, R_X \, sin \, \theta$, the coordinate of the center
of this circle is at a distance $y = R_X \, cos \, \theta$, i.e. symmetric
about the origin of the $x, ~y, ~z$ system (i.e. the location of the 
black hole) and the intersection of the observer -- X-ray source
line on the $x,~y$ plane. One can easily see that for the above value
of the lag $D$, the isodelay surfaces are tangent to the plane 
$ z =  y \, cot \, \theta$ at the point $x = 0, ~y  = R_X \, 
cos \, \theta, ~z  = R_X \, cos^2\theta/sin \, \theta$.

In order to compute the equation of the intersection on the plane of 
the accretion disk, the above equation has to be expressed in terms
of the coordinates $x^{\prime \prime},
~y^{\prime \prime}, ~z^{\prime \prime} $ on the plane $ z =  y \, 
cot \, \theta$. These are related to $x, ~y, ~z$ by the relation 
$x^{\prime \prime}
= x$, $ y^{\prime \prime\, 2} = y^2 + z^2 = y^2 + y^2 \, cot^2 \theta =
y^2/sin^2 \theta$, $z^{\prime \prime} = 0$; i.e. one has to  project 
this circle onto the plane of the accretion disk. This, as expected, 
yields the equation of an ellipse, namely
\begin{equation}
\left[y^{\prime \prime}\, sin \, \theta -  (D \, cot \, \theta - R_X \, 
cos \, \theta)\right]^2  +  x^{\prime \prime 2} =
\frac{D}{sin \, \theta} \left( \frac{D}{sin \, \theta} - 2 R_X \right)
\end{equation}
Dropping the primes and defining polar coordinates on this plane, 
$x = R \, sin\,\phi$, $y = R \, cos\,\phi$, 
one obtains easily, solving for $D$, 
\begin{equation}
D = \sqrt{R^2 + R_X^2} - R \, cos \, \theta \, cos\,\phi + R_X
\, sin \, \theta ~.
\label{rokaki}
\end{equation}
\noindent This is the equation for the delays used by Rokaki \& Magnan
(1992), with the difference that their angle $\theta$ denotes the 
colattitude (inclination angle) rather than the lattitude of the 
observer with respect to the accretion disk, given in our notation.

One can now compute, by setting the value of the angle $\phi$ in 
Eq. (\ref{rokaki}) to $\phi = 0,~ \pi$, the range of values of 
the delay $D$ for which a ring of given radius $R$ will be 
illuminated by an instantaneous flash of X-rays emitted by the
source $S$ at $t=0$. These are
\begin{equation}
D_l = \sqrt{R^2 + R_X^2} - R \, cos \, \theta + R_X \, sin \, \theta
~~~~, ~~~~D_t = \sqrt{R^2 + R_X^2} + R \, cos \, \theta + R_X \, 
sin \, \theta
\end{equation}
where the subscripts $l$ and $t$ refer to the leading and trailing
times. Then, the entire interval of the illumination of a
ring of radius $R$ on the plane of the accretion disk is 
$\vert D_l - D_t \vert = 2 \, R \, cos \, \theta$,  while the 
position at which the isodelay surface of lag $D$ touches the 
plane of the accretion disk is given by the projection of the 
circle of Eq. (\ref{circle}) onto this plane, i.e. at the point
of radius $R = R_X \,cos \, \theta / sin  \, \theta = R_X \, 
cot \, \theta$. 

Using Eq. (\ref{rokaki}) one can 
now compute the area of overlap between these two families of 
curves. The simplest way to do this is to form the cross product
of the tangent vectors of these two families of curves at the 
points of their intersection $(x_i,~y_i)$. These vectors are
$(dx_i/dD,~dy_i/dD)$ and $(dx_i/dR,~dy_i/dR)$. After a considerable
amount of algebra, performed  with the use of {\sl Mathematica}, 
the expression for the overlap area reduces to the following simple
expression
\begin{equation}
A(D, R) = \frac{2 \, R}{\sqrt{(D - D_l)(D_t -D)}}
\end{equation}
with the values of $D_l, ~ D_t$ given above. 

One can now integrate the above area over all $D$ between $D_l$ and 
$D_t$. This integration yields $ 2 \, \pi \, R$ indicating that the
Area function as given above is properly normalized. Therefore, the
emitting area as a function of the time lag $\tau$ is
\begin{equation}
A(\tau) = \cases{ 2 \, R/\sqrt{(\tau - \tau_l)(\tau_t -\tau)}
 & if $\tau_l < \tau < \tau_t$ \cr
0 & otherwise \cr}~.
\end{equation}
where $\tau_l = D_l/c$ and $\tau_t = D_t/c$.

\begin{figure*}[H]
\centerline{\psfig{file=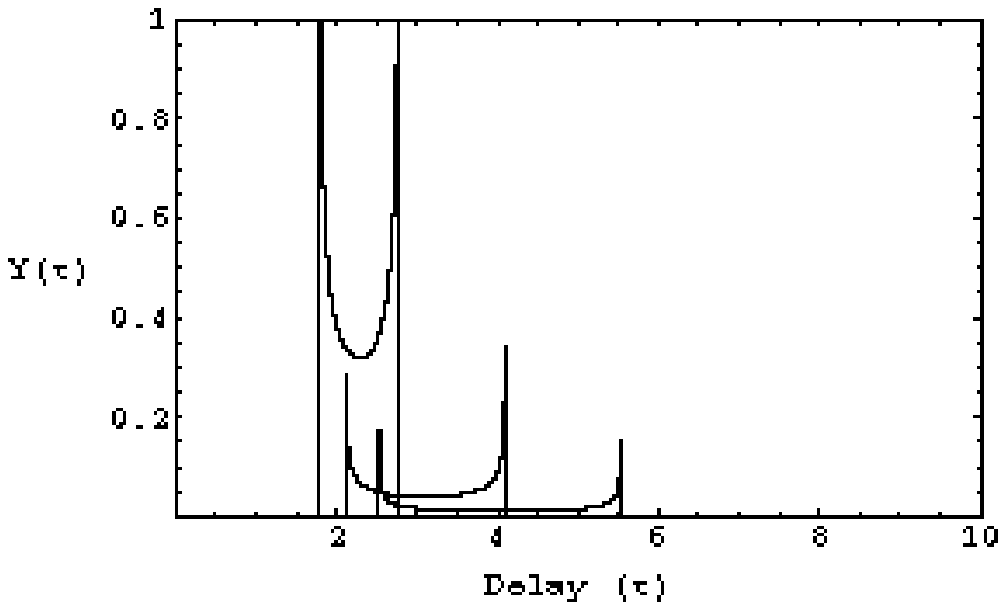,width=.60\textwidth}}
\caption{The area of intersection of constant $R$ annuli with 
the constant $D$ ellipses divided by $R^3$ for $\theta = \pi/3$ 
and $R = 1, 2, 3 \; R_X$. The units of delay are $R_X/c$. } 
\label{horns}
\end{figure*}

In Figure \ref{horns} we present the ratio $Y(\tau) =A(\tau)/R^3$ 
of the overlap 
function $A(\tau)$ divided by the cube of the radius of the disk 
as a function of the delay $\tau$ in units of $R/c$, for $\theta
= \pi/3$ and assuming $R_X =1$. This is a 
quantity of interest as it indicates the reprocessed flux,
integrated over all frequencies, contributed at a given lag $\tau$ 
by  the appropriate radii, $R = 1, 2, 3$ in  units of $R_X$. 
The figure  makes  apparent the rapid decrease of the reprocessed
radiation with radius and also the increasing range of delays $\tau$
which contribute to the emission at a given radius with the increase
of the radius.  For the  given geometry, most of the 
reprocessing is effected by the shortest radii $R \sim R_X$.
Also apparent in the figure are the (integrable) singularities
associated with the instants of intersection of the constant temperature 
circles with the constant delay ellipses. 

The response function $F(\l,\tau)$ at a wavelength \ls and lag $\tau$
will be the flux contributed at a given wavelength \ls by all 
radii of  the disk which emit at the given lag $\tau$. 
Assuming  that the reprocessed  radiation is emitted locally with a 
black body spectrum,  the response function has the form 

\begin{equation}
F(\nu,\tau) = \int_{r_{\rm min}}^{r_{\rm max}} B_{\nu} [T(R)] A(\tau) 
\; dR 
\label{response}
\end{equation}

\noindent
where $B_{\nu} [T(R)]$ is the Planck function of temperature $T(R)$. 
The limits of integration 
$r_{\rm min}$ and $r_{\rm max}$ are the minimum and maximum radii 
contributing to a given lag and are computed using Eq. (\ref{rokaki})
by setting $c \, \tau = D_l$ and $c \, \tau = D_t$. 

The temperature 
$T(R)$ is calculated assuming that a fraction $1 - \cal{A}$ 
($\cal{A}$ $\ll 1$ is the albedo of the disk) of the X-ray flux incident 
at a given point on the disk is thermalized and re emitted in 
black body form, i.e.

\begin{equation}
\sigma \, T(R)^4 = \frac{L_X \, (1 - \cal{A})}{4\, \pi \,(R_X^2 + R^2)} 
\frac{R_X}{(R_X^2 + R^2)^{1/2}} 
\end{equation}
or
\begin{equation}
T(R) = \left[\frac{L_X\, (1 - {\cal{A}}) \, R_X}{4 \,\pi \, \sigma} 
\right]^{1/4} \frac{1}{(R_X^2 + R^2)^{3/8}}
\end{equation}

If the source of X-rays is not point-like  but extended, a different 
relation between the flux and the distance to the disk  would result;
Rokaki \& Magnan (1992), for example, have used the expression for a 
spherical source of radius $R_X$. This would lead to a slightly different 
relation  between $T$ and $R$ for $R \sim R_X$. However, for large $R$ 
the above  relation is essentially correct.

\section{Reprocessing the X-ray Flux}

Having obtained the expression  for the response of the disk to 
X-ray illumination as a function of the frequency of the reprocessed
radiation and the lag $\tau$, one can now proceed to the 
computation of the time dependence of the reprocessed radiation at 
a given wavelength \l.
If $S_X(t)$ is the light curve of the observed X-ray radiation, 
then, the reprocessed emission at a wavelength \ls  as a function of 
time, $f_{\l}(t)$,  will be given by
\begin{eqnarray}
f_{\l}(t) &=& \int d \tau \, F(\l,\tau) \, S_X(t - \tau) \\
           &=& \int_0^{\infty} d R \, B_{\l} [T(R)] \, 
\int_{\tau_l}^{\tau_t} d \tau \, A(\tau) \, S_X(t - \tau)
\label{lc}
\end{eqnarray}

It is apparent from the discussion of the previous section that the 
quantity which sets the scale of the lag is the distance
of the X-ray  source  from the  disk, or alternatively, for
extended  sources the size of the X-ray emitting region, $R_X$. 
We shall assume in the rest of this note, in order to  fix our units,
that this distance is equal to the \sc radius of the black hole. Since 
the scale of the temperature of the reprocessed radiation for the given, 
observed luminosity of NGC 7469 at a given radius $R$ is scaled  by the 
value  of $R_X$ measured in cm, should $R_X$ be larger than the \sc radius
by a given factor, this could simply be reabsorbed in the value of the 
mass of the black hole, which would have to be decreased by the 
same factor.

\begin{figure*}[H]
\centerline{\psfig{file=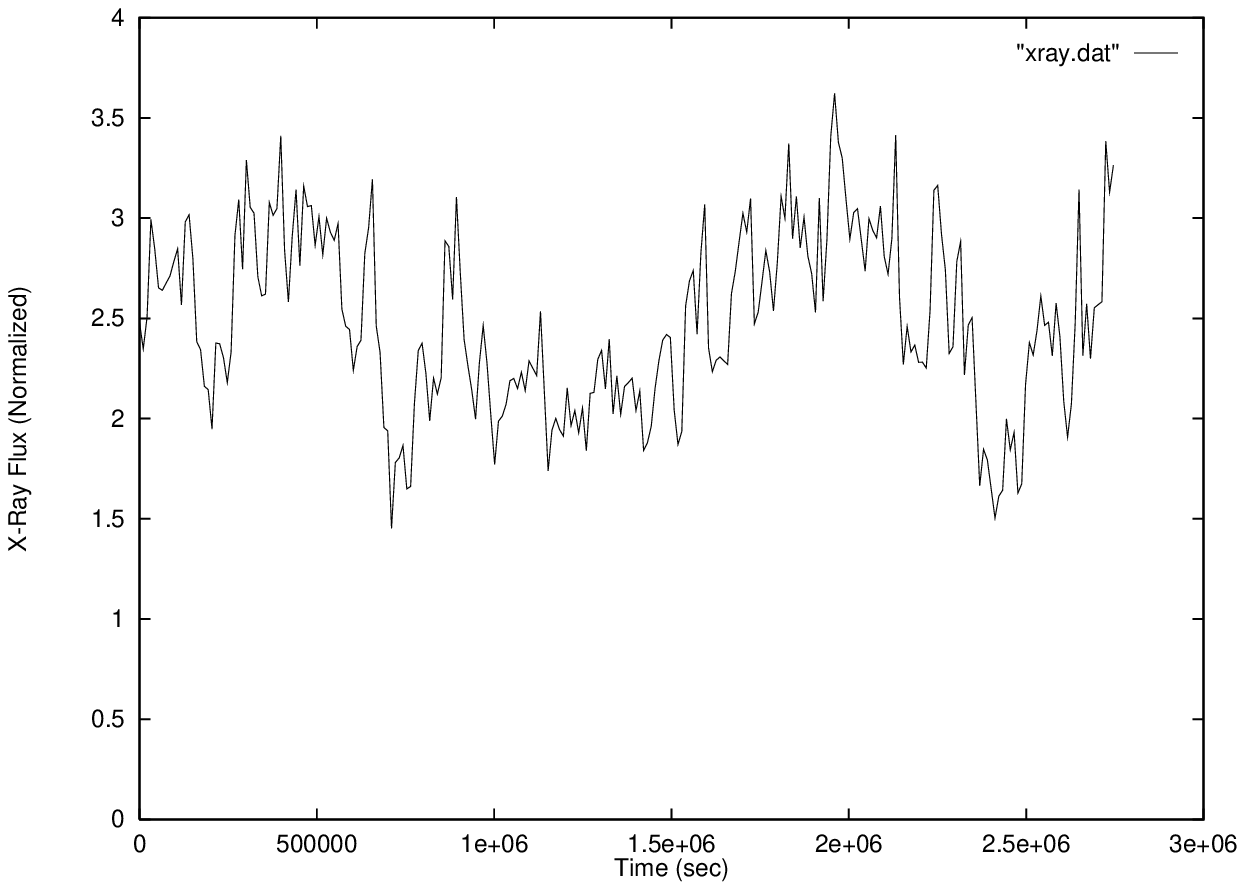,width=.65\textwidth}}
\caption{The normalized X-ray light curve in the 2-4 keV range. } 
\label{xray}
\end{figure*}

Using as input the observed X-ray flux as a function of time, we 
employed Eq. (\ref{lc}) to compute the corresponding variations 
of the reprocessed flux at a number of wavelengths \l, for a set 
of values for  the black hole mass $M$ and for two values of the 
latitude angle of the observer with respect to the plane of the 
disk $\theta$. The (normalized) X-ray light curve in the 2-4 keV 
range we used is given in Figure \ref{xray}. This light curve 
consists of 256 equally spaced points, each representing the average 
flux in an interval of 10,768 seconds, provided to us by P. Nandra. 
The flux was obtained by 
interpolation of the real light curve at the equally spaced 
intervals given in the figure. The details of these observations 
are given in Nandra et al. (1998). It is worth noting that in 
the first 10 days of the observing campaign the sampling interval
was roughly half of that given in the figure (once every 90 min., or 
5,400 s), while for the remaining of the campaign it was once every 180 
minutes or 10,800 seconds. The values of the black hole mass used 
in our calculations were $M = 10^7, \, 10^8, \, 10^9 \, 
{\rm M}_{\odot}$, while for the angle $\theta$ we used the values
$\theta = \pi/3, \pi/10$.

\begin{figure*}[H]
\centerline{\psfig{file=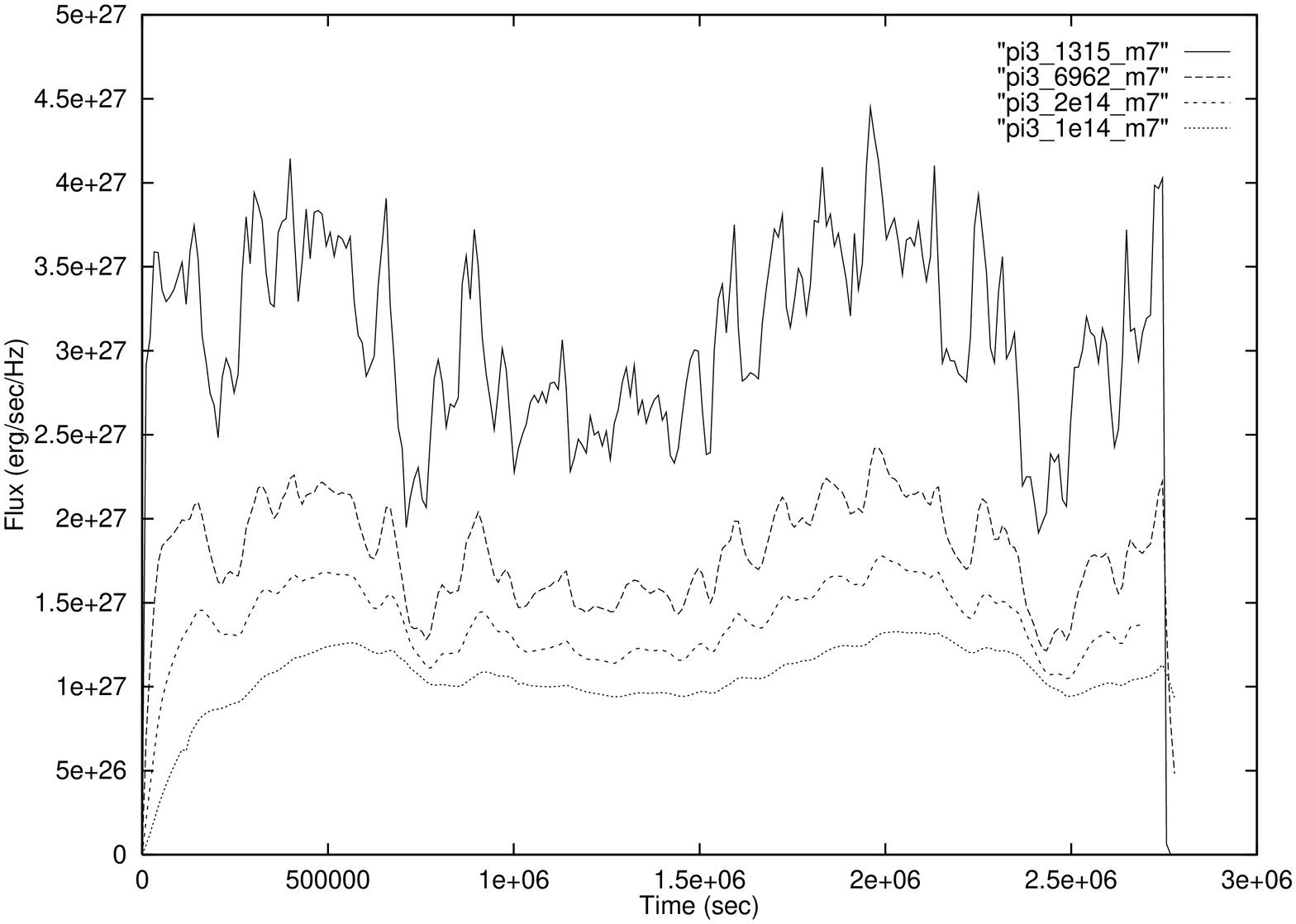,width=.65\textwidth,angle=-90}}
\caption{The reprocessed radiation light curves for a set of wavelengths
(\l\l 1315, 6962, 15000 and 30000 \AA)  for $M = 10^7 {\rm M}_{\odot}$ and 
$\theta = \pi/3$. } 
\label{m7lcurv}
\end{figure*}

Figure \ref{m7lcurv} exhibits a representative set of such light 
curves corresponding 
to $M = 10^7 \, {\rm M}_{\odot}$ and $\theta = \pi/3)$ for four different
wavelengths, i.e. \l\l 1315\AA, 6962\AA, 15000\AA~ ($2 \times 10^{14}$ Hz) 
and 30000\AA~  ($10^{14}$ Hz). The first two wavelengths were chosen
to match those at which the UV and optical observations were made, while
the last two in order to explore whether model light curves at 
different  wavelengths could reproduce the general characteristics 
of the observed light curves. The general trend apparent in this figure
is the ``smoothing" of light curves with increasing wavelength, this 
``smoothing" becoming apparent already at 
\l 6962 \AA. Such a trend is not surprising, as the contribution to the 
flux at these longer wavelengths comes from increasingly larger radii.
In addition to the smoothing of the sharpest features associated
with the X-ray  light curve, the  RMS variability also decreases
significantly with increasing wavelength from $\simeq 50\%$ at
\l 1315 \AA, to $\sim 33\%$ at \l 6962 \AA, to $\sim  25\%$ at
\l 15000 \AA~ ($2 \times 10^{14}$ Hz), since an increasing range in 
the lag contributes to the emission at a given wavelength.

\begin{figure*}[H]
\centerline{\psfig{file=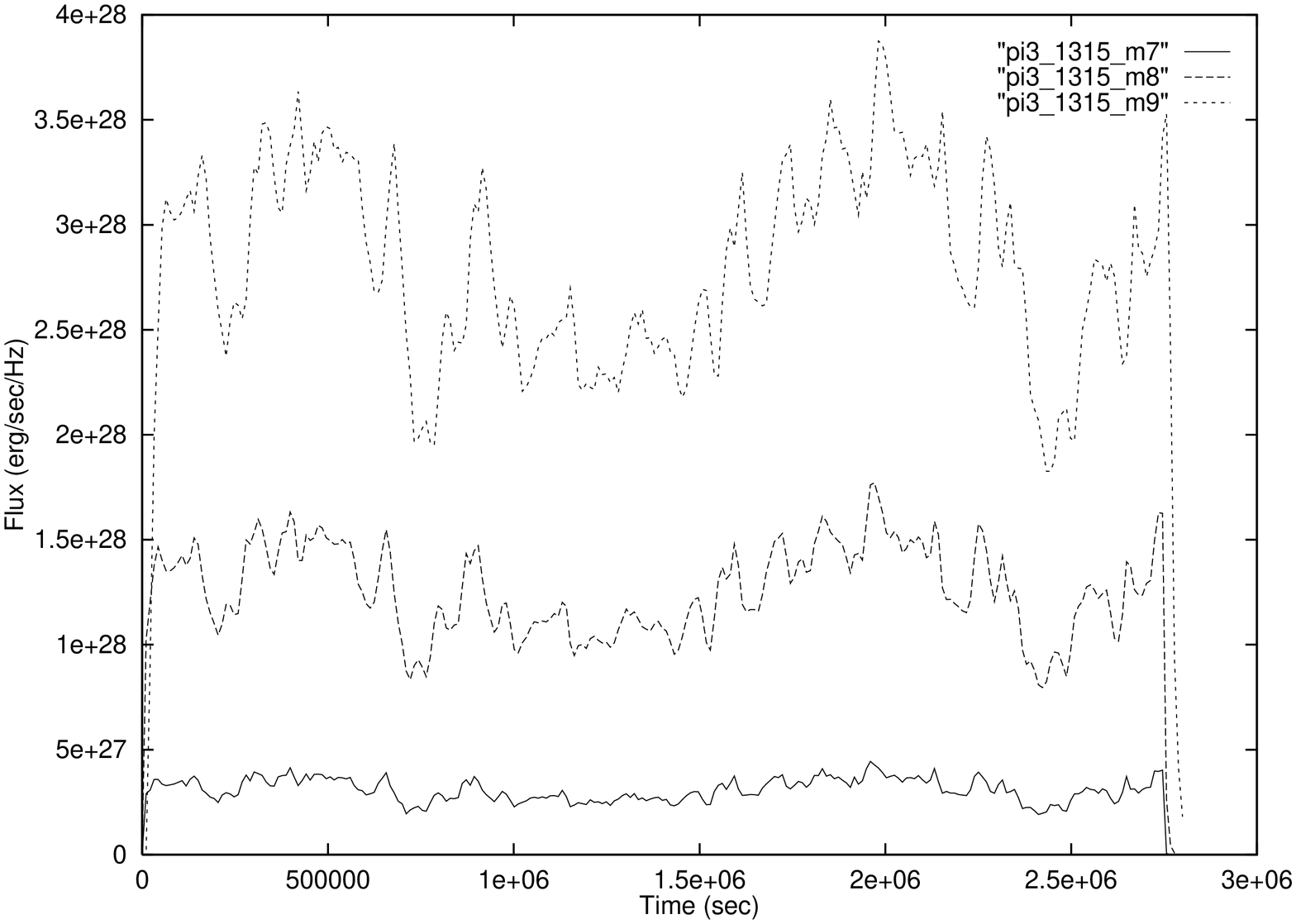,width=.6\textwidth,angle=-90}}
\caption{The reprocessed radiation light curves at wavelength $\l = 
1315$ \AA, for three values of the black hole mass $M = 10^7, \, 
10^8, \, 10^9 \, {\rm M}_{\odot}$ and  $\theta = \pi/3$. } 
\label{m789lcurv}
\end{figure*}

\begin{figure*}[H]
\centerline{\psfig{file=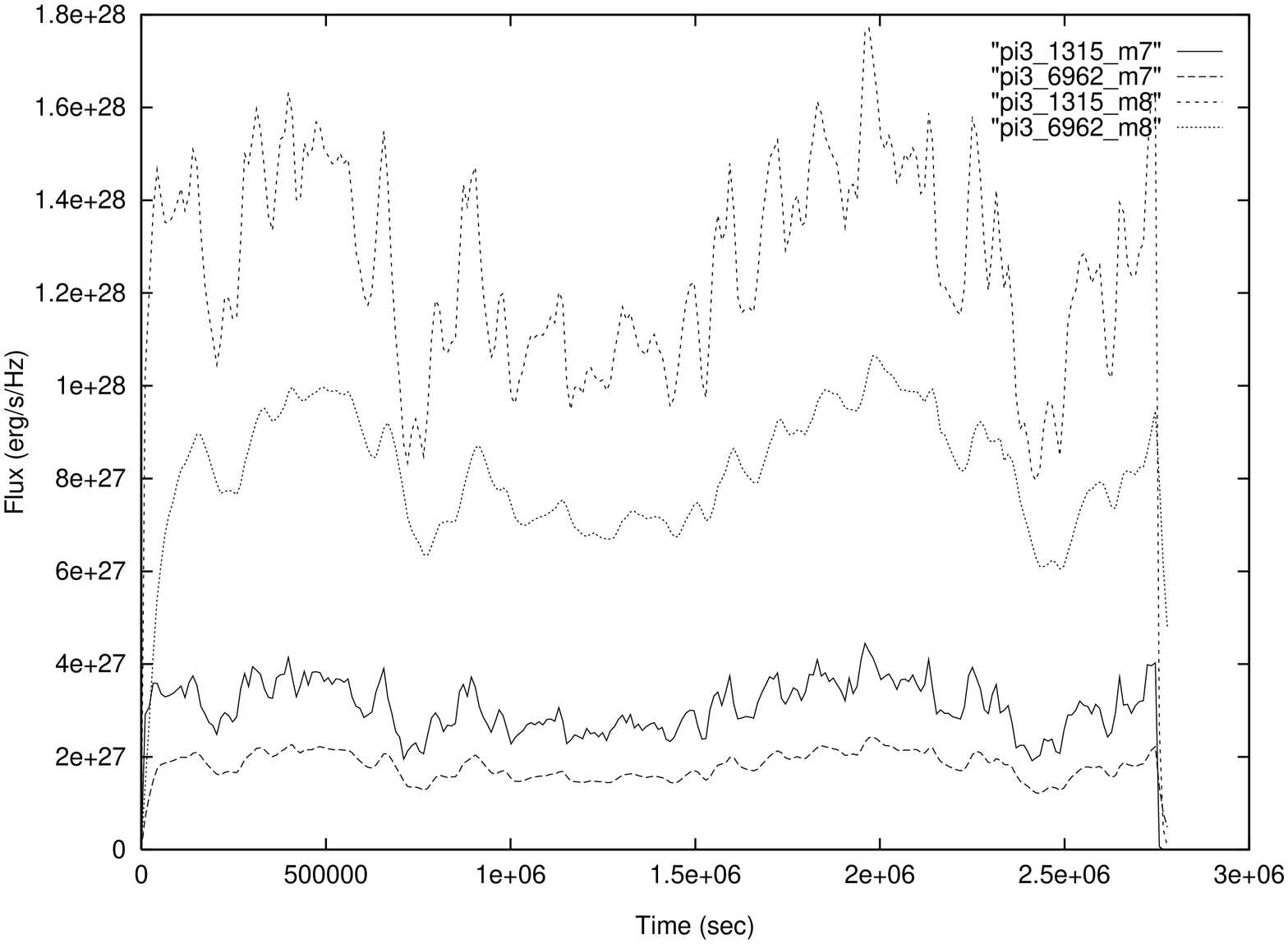,width=.6\textwidth,angle=-90}}
\caption{The reprocessed radiation light curves at two wavelengths $\l \l 
1315, \, 6962$ \AA, for two values of the black hole mass $M = 10^7, \, 
10^8 \, {\rm M}_{\odot}$ and  $\theta = \pi/3$. } 
\label{m7m8_2lcv}
\end{figure*}

Figure \ref{m789lcurv} exhibits the effect of the black hole mass $M$
(or the height of the location of the X-ray source above the disk)
on the light curves of the reprocessed radiation, where the light
curves at a specific wavelength (\l 1315 \AA) are shown for three different
values of the black hole mass $(M = 10^7, \, 10^8, \, 10^9 \, 
{\rm M}_{\odot})$. The most obvious feature in this figure is the 
increase of the flux with the increase of the black hole mass. 
This fact can be understood as follows: for a given constant 
reprocessed luminosity, an increase in the mass corresponds to 
a decrease in the corresponding disk effective temperature; since the 
1315  \AA~  wavelength is in the Rayleigh--Jeans part of the 
corresponding spectra, a smaller temperature requires, in this 
regime of the spectrum, a larger flux in order that the same luminosity
be radiated away. Increase in the black hole mass leads also to a 
certain amount of smoothing and shifts in the corresponding light 
curves, which become most prominent at the largest values of this 
parameter (see also the figures of the autocorrelation and cross c
correlation functions). These trends are expected, as a larger value 
for the black hole mass corresponds to a larger distance of the 
X-ray source from the reprocessing disk plane. In this respect, 
one should note that the increase of the mass $M$ from $10^7$ to 
$10^8$ M$_{\odot}$ causes very little additional smoothing at 
\l1315 \AA. The flux in this wavelength is emitted 
by a region of extent $\sim R_X$, which, in both these cases, 
is smaller than the light crossing distance associated with the 
X-ray light curve sampling interval ($\simeq 10^4$ s), i.e. 
$R_X \lsim 3 \times  10^{14}$ cm. The smoothing and shifting trends 
become more pronounced at longer wavelengths as it becomes apparent in
Figure \ref{m7m8_2lcv}, where the reprocessed light curves are shown for 
two values of the wavelength (\l\l = 1315, 6962 \AA) and two values of 
the black hole mass $(M = 10^7, 10^8 \, {\rm M}_{\odot})$.

\begin{figure*}[H]
\centerline{\psfig{file=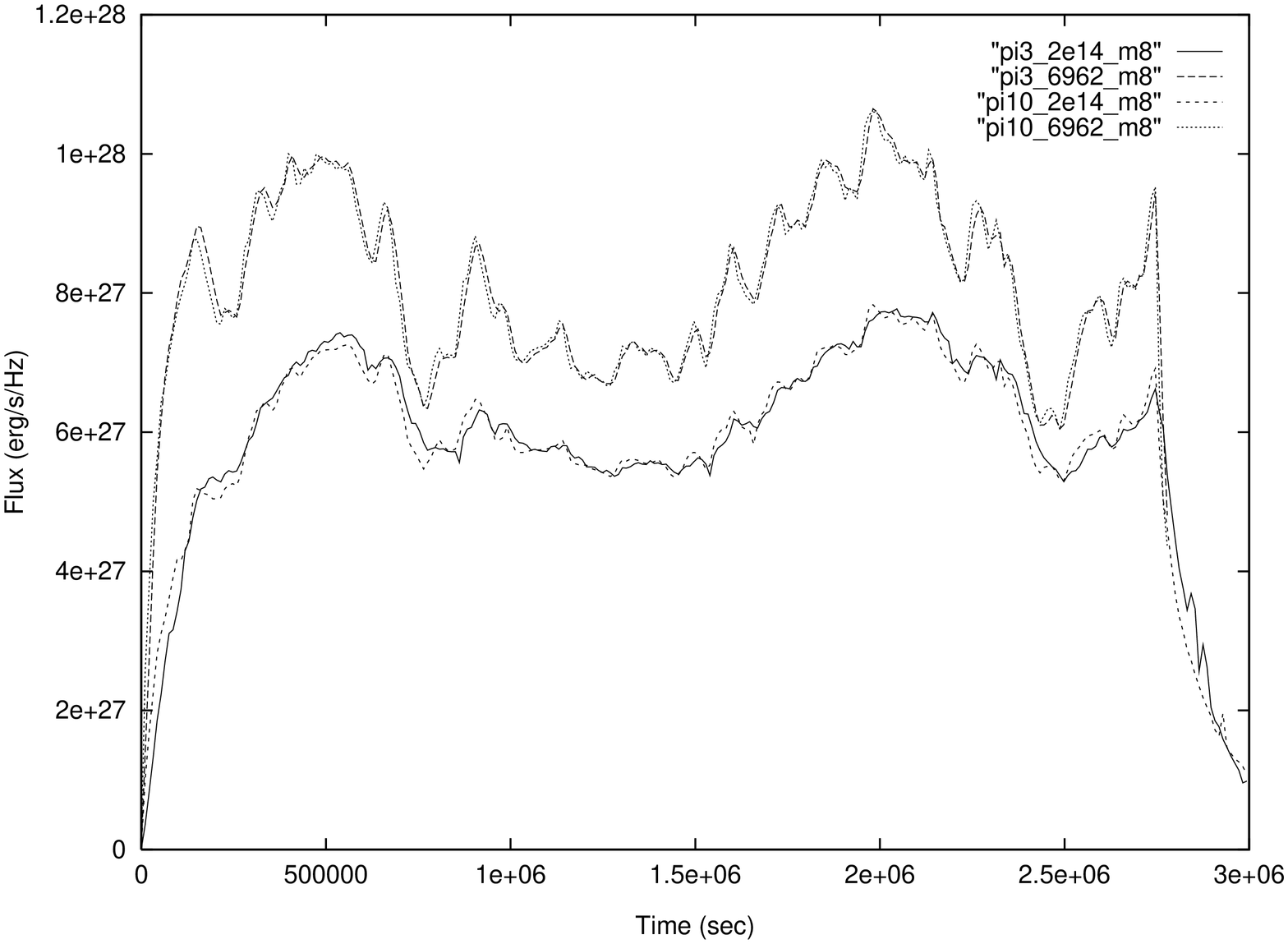,width=.6\textwidth,angle=-90}}
\caption{The reprocessed radiation light curves at two wavelengths $\l \l 
6962, \, 15000$ \AA, for $M = 10^8 \, {\rm M}_{\odot}$ and two values
of the colatitude  $\theta = \pi/3, \pi/10$. } 
\label{pi3pi10}
\end{figure*}

The effects of the inclination angle on the resulting light curves 
are shown in figure \ref{pi3pi10} which exhibits these  light curves at
\l\l 6962, 15000 \AA~ for two different values of the observer's
latitude $(\theta = \pi/3, \pi/10)$. One can see that the effect
of the latitude on the light curve does increase with increasing
wavelength: It is barely discernible at \l 6962 \AA~ (it is even
smaller at \l 1315 \AA), but it is easily observable (though still
small) at \l 15000 \AA. Qualitatively, the larger inclination angle 
light curve $(\theta = \pi/10)$   preserves slightly more of the high 
frequency content of the X-ray light curve than does the lower 
inclination one ($\theta = \pi/3$) and it certainly has a faster
``turn--on" phase. The reason for that is that
in the former case part of the reprocessing takes place 
almost along the observer's line of sight and thus does not 
get ``washed-out" by time-of-travel across the source effects. 

To cast the arguments given above concerning the ``smoothing" and 
``shifting" of the model light curves with increasing  
black hole mass and wavelength in more quantitative form,  we have
computed the autocorrelation  (ACF) functions of our model light 
curves (a sample of which is given in the figures above) as well
as their cross correlation (CCF) functions with the input X-rays. 
We have searched the black hole mass $M$ and angle $\theta$ parameter
space with the intent of comparing these ACFs and CCFs to those 
of NGC 7469 given in Nandra et al (1998). A slight technical 
detail should be mentioned here: the model light curves we 
produce using as input that of the X-rays have 
a ``turn--on" phase; had this section of the light curve been
included in the computation of the ACF and CCF, it would lead to
unphysical forms for these functions. For instance, this ``turn--on" 
feature would lead to an ACF for the \l 1315 \AA~ light curve which 
is narrower than that of the X-ray light curve, an unphysical 
result since the latter signal can at best track the former exactly. 
To avoid such unphysical effects, in computing the ACF and CCF, 
we have excluded this ``turn--on" section from the corresponding 
light curves. For the computation of the CCFs in particular, we 
have excluded at the same time an interval of equal length from 
the X-ray light curve too.

\begin{figure*}[H]
\centerline{\psfig{file=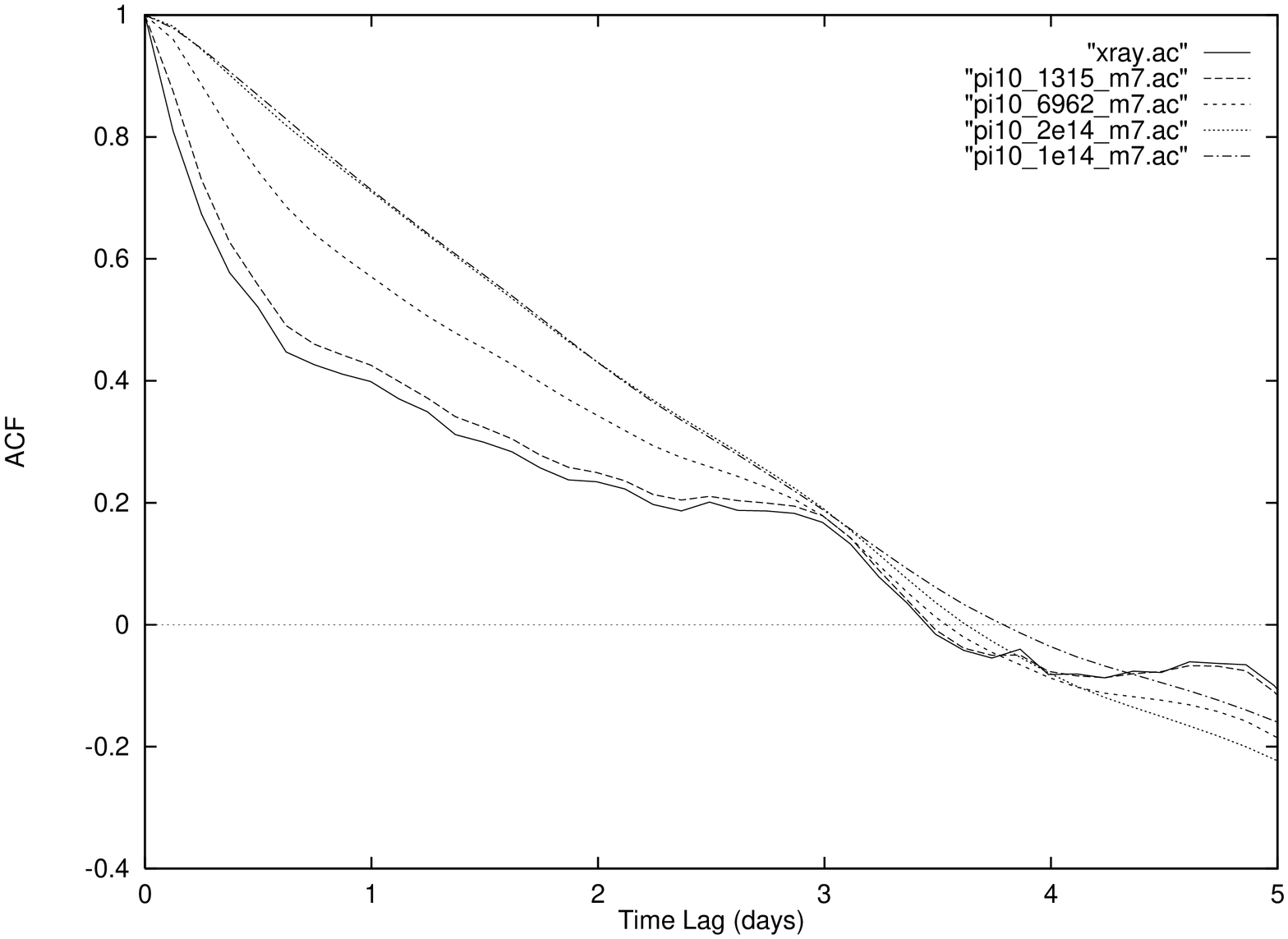,width=.6\textwidth,angle=-90}}
\caption{The autocorrelation function (ACF) of the X-rays and the 
computed light curves at wavelengths $\l \l 1315, \, 6962, \, 15000,
\, 30000$ \AA, for $M = 10^7 \, {\rm M}_{\odot}$ and  $\theta = \pi/10$. } 
\label{acfm7}
\end{figure*}

\begin{figure*}[H]
\centerline{\psfig{file=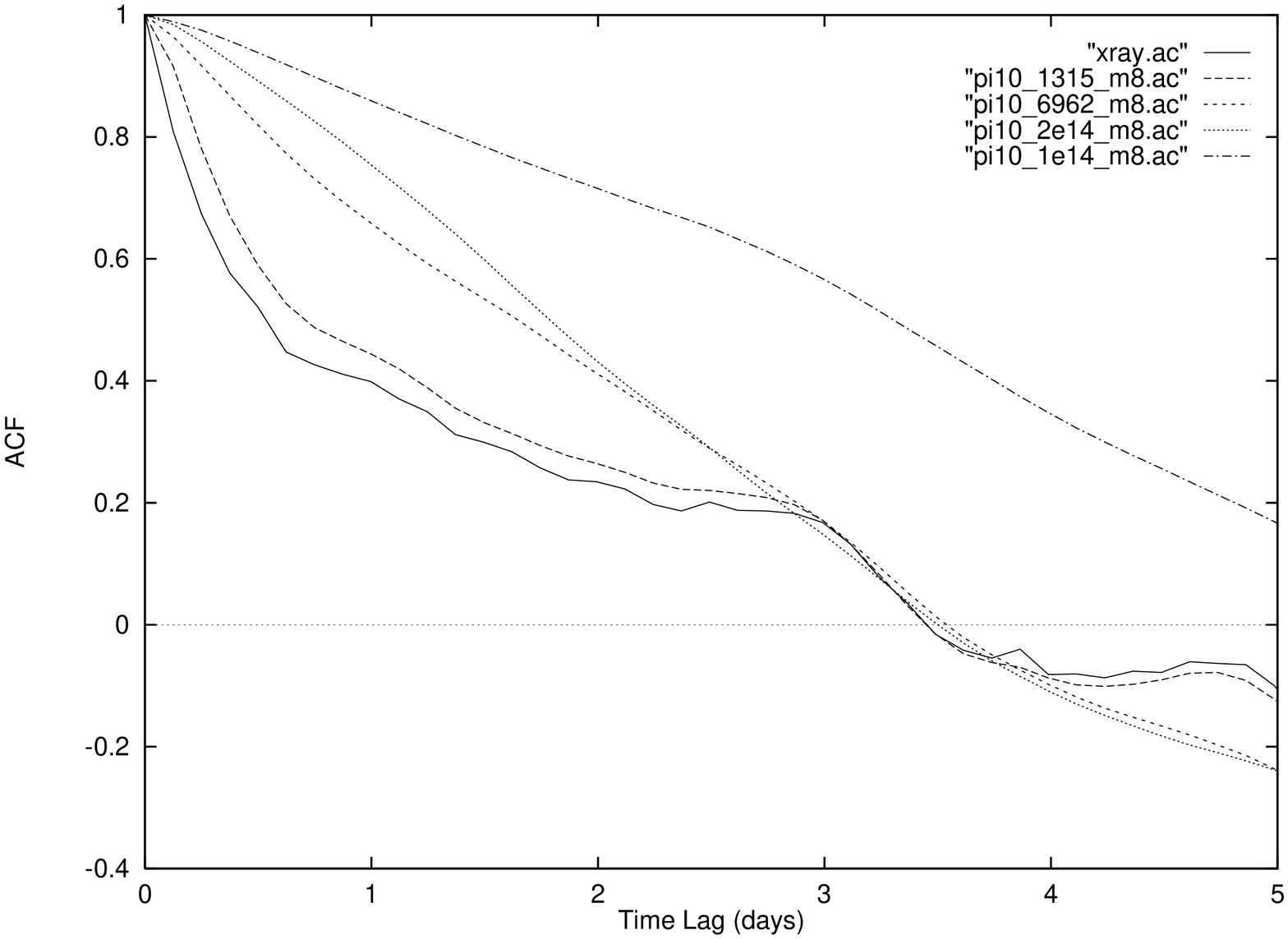,width=.6\textwidth,angle=-90}}
\caption{Same as Figure \ref{acfm7} but for $M = 10^8, \, 
{\rm M}_{\odot}$.} 
\label{acfm8}
\end{figure*}

The ACFs are given in figures \ref{acfm7} and \ref{acfm8} for two 
different values of the black hole mass ($M = 10^7, 10^8 \, 
{\rm M}_{\odot})$, $\theta = \pi/10$  and for the four
different wavelengths \l \l \, 1315\AA, 6962\AA, 15000\AA~ ($2 \cdot 
10^{14}$ Hz), 30000\AA~ ($10^{14}$ Hz) along  with the ACF of the 
X-ray light curve itself depicted by the solid curve. These figures
make more quantitative the statements made above concerning the 
``smoothing" of light curves with wavelength and black hole mass:
The ACF of the \l 1315 \AA~ light curve is almost identical to that
of the X-rays and it changes very little with increasing the black
hole mass from $10^7$  to $10^8$ M$_{\odot}$. The ACFs become broader
with increasing both the black hole mass and the wavelength of 
the corresponding emission, indicating progressively ``smoother" 
light curves with changes in these parameters, in agreement with the 
above discussion. 

\begin{figure*}[H]
\centerline{\psfig{file=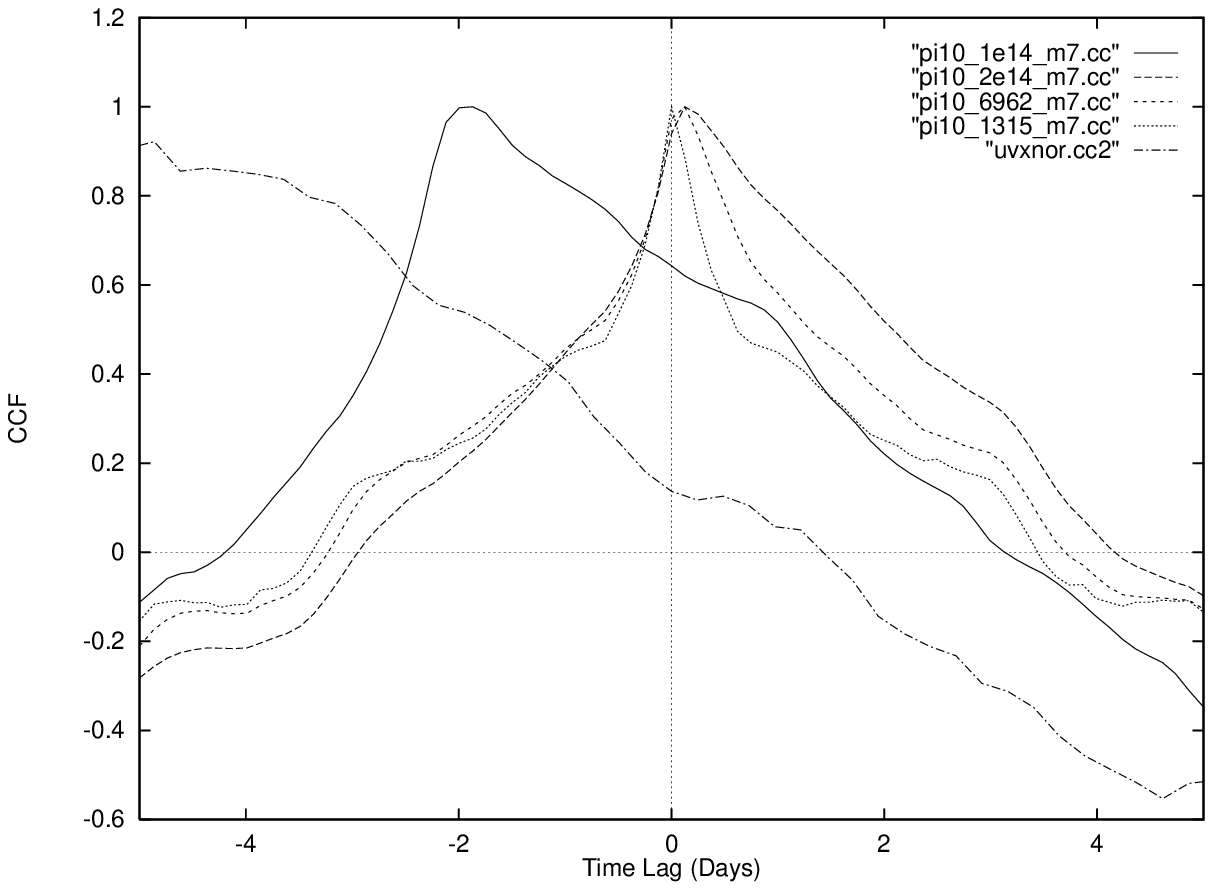,width=.6\textwidth}}
\caption{The cross correlation function between the X-ray and the 
model light curves at \l \l 1315,  6962,  15000 and 30000 \AA, for 
$M = 10^7 \,  {\rm M}_{\odot}$ and  $\theta =  \pi/10$. The cross 
correlation function between the {\sl observed} X-ray and UV light curves
(dashed-dot curve) is also shown for comparison. } 
\label{ccfm7}
\end{figure*}

\begin{figure*}[H]
\centerline{\psfig{file=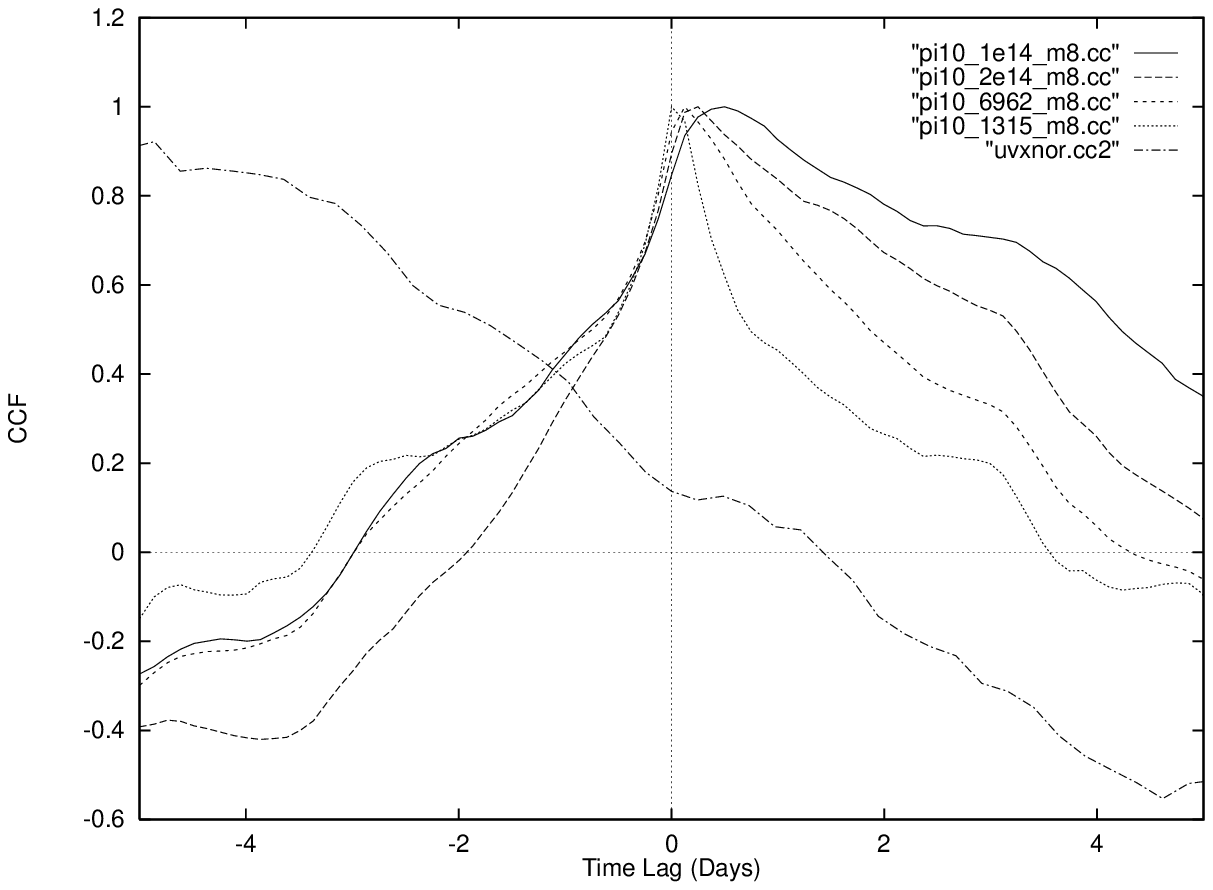,width=.6\textwidth}}
\caption{Same as in Figure \ref{ccfm7} for $M = 10^8, \, 
{\rm M}_{\odot}$. } 
\label{ccfm8}
\end{figure*}

The  shift in time of the emission due to X-ray reprosessing relative 
to the incident X-rays themselves is given by their corresponding cross 
correlation functions (CCF). These are shown in Figures \ref{ccfm7} 
and \ref{ccfm8} respectively along with the observed value of the
X-ray--UV CCF (Nandra et al. 1999), for the same values of the parameters 
used to produce the ACFs of Figures \ref{acfm7} and \ref{acfm8}. 
The trends are similar to those of the ACFs: Increasing black hole 
masses and wavelengths lead to longer lags between the X-rays and 
the reprocessed radiation, again in agreement with the qualitative
arguments made earlier. An exception to this rule is CCF of
the light curve of \l 30000 \AA~ for $M = 10^7$ M$_{\odot}$ and 
$\theta = \pi/10$. The CCF these values of the parameters exhibits
a behavior not unlike that corresponding to the CCF between the 
X-rays and the UV emission at \l 1315 \AA~ in the data of Nandra 
et al. (1998). We believe that this particular feature is related 
to the specific form of the X-ray light curve. As it is apparent it
disappears for a larger value of the black hole mass (see Figure 
\ref{ccfm8}).

\begin{figure*}[H]
\centerline{\psfig{file=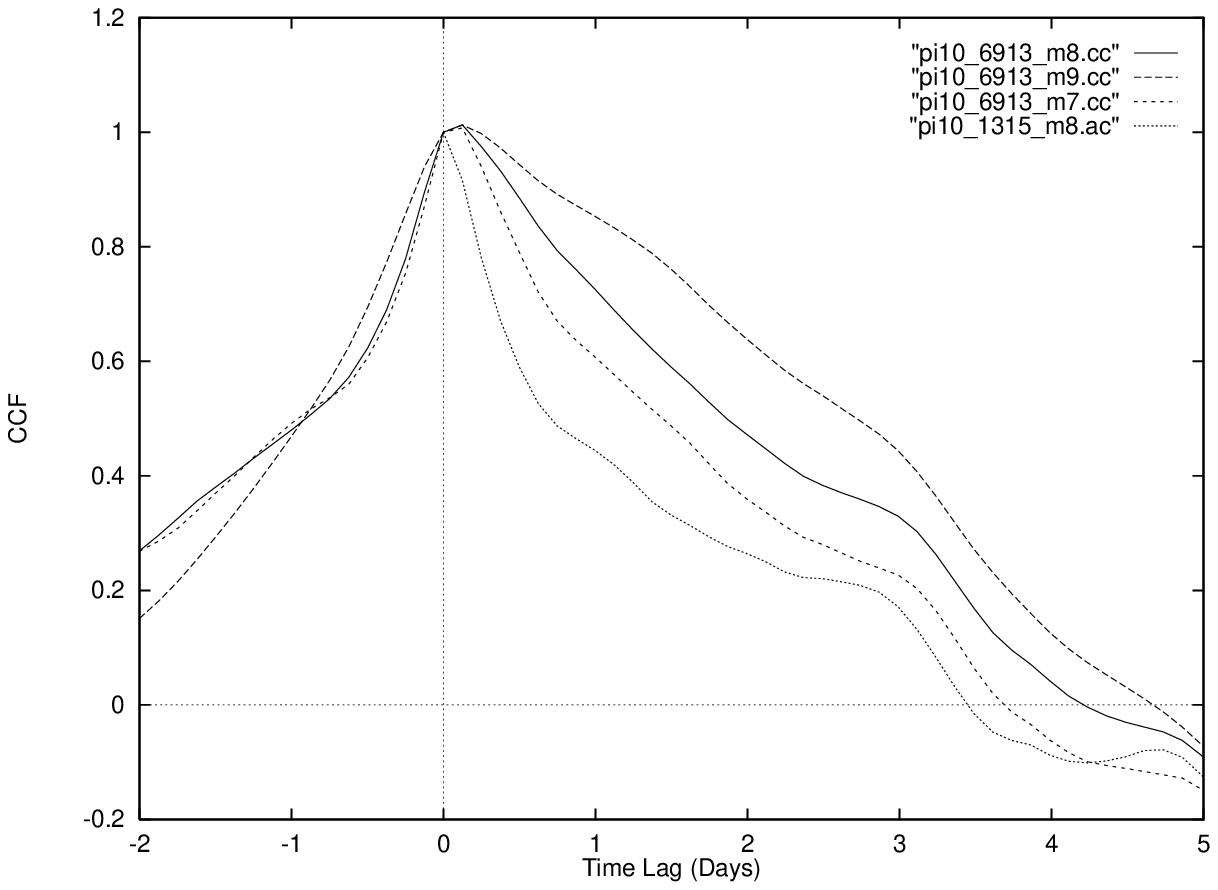,width=.6\textwidth}}
\caption{The cross correlation function between the model light curves
at $\l 1315$ and $\l \, 6962$ \AA, for three values of 
the black hole mass $M = 10^7, \, 10^8, \, 10^9 \, {\rm M}_{\odot}$ and  
$\theta =  \pi/10$, assuming that the variability in these bands is 
driven by the varying X-ray flux. The value of the autocorrelation 
function for \l 1315 \AA~ and $M = 10^8$ is also shown for comparison. } 
\label{relccf}
\end{figure*}

\begin{figure*}[H]
\centerline{\psfig{file=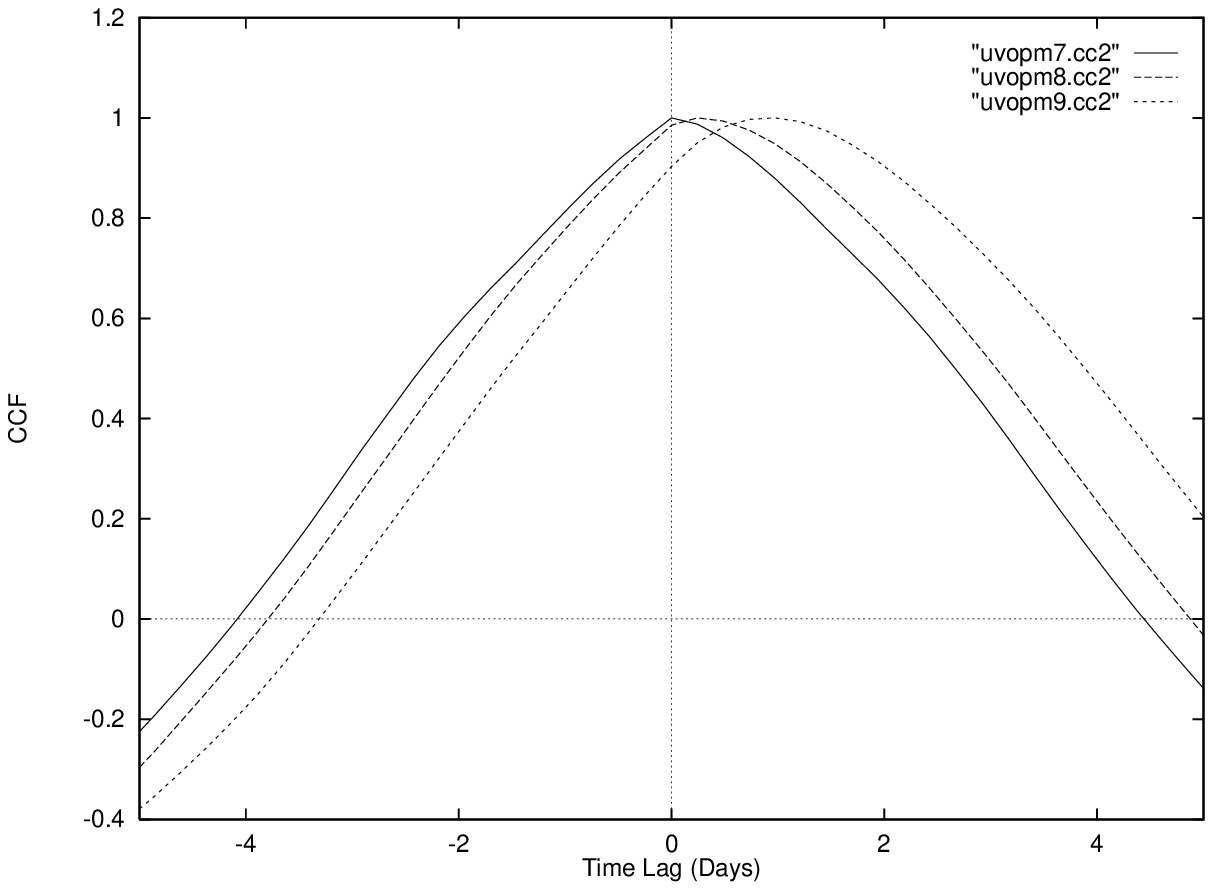,width=.6\textwidth}}
\caption{The cross correlation function between the model light curves
at $\l 1315$ and $\l \, 6962$ \AA, for three values of the black 
hole mass $M = 10^7, \, 10^8, \, 10^9 \, {\rm M}_{\odot}$ and  
$\theta =  \pi/3$, assuming that the variability in 
these bands is driven by an unseen spectral component (EUV) with 
luminosity and light curve identical to that of the observed UV emission.} 
\label{relccf2}
\end{figure*}

Finally, in order to exhibit our results in a manner similar to that of 
Collier et al. (1998), who discussed in detail the UV -- optical 
continuum lags of NGC 7469, and cast them in the parlance of accretion
disk physics, we have produced the cross correlation of our model light
curves at \l6962 \AA~ with those at \l1315 \AA. We have done that
in two ways: (a)First we assumed, as done so far, that the variability
in both these bands is driven by the X-rays. The resulting 
cross correlations functions are given in 
Figure \ref{relccf} for three values of the black hole mass $M = 
10^7, \, 10^8, \, 10^9 \; {\rm M}_{\odot}$ and $\theta = \pi/10$. 
As expected, since either light curve tracks very closely that of
the X-rays, their cross correlation peaks at time scales associated
with the X-ray variability. The observed lags are therefore too 
long to be accounted by the geometric arrangement. 
(b) In the spirit of the Rokaki \& Magnan (1992) treatement
(and at the referee's suggestion) we have assumed that the UV - optical
variability is driven by an unseen spectral component other than
the  X-rays (EUV ?), with a light curve identical to that of the 
observed UV emission. We then computed the resulting light curves 
at \l1315 \AA~ and \l6962 \AA~ and their cross correlation functions 
which are shown in Figure \ref{relccf2}, for  the same values of the 
black hole mass as before, namely $M = 
10^7, \, 10^8, \, 10^9 \; {\rm M}_{\odot}$ and for $\theta = \pi/3$.
The smoother variations of the UV band lead to a broader CCF, 
which for $M = 10^9 \; {\rm M}_{\odot}$ leads to a lag of about 
1 day, consistent with those given in Collier et al. (1998). 
However, in this specific value of $M$, the UV light curve at \l1315 
\AA~ is visibly smoother than the observed light curve at the
same wavelength. For the smaller values of $M$ the incident
and reprocessed UV light curves are in fact almost identical but
in this case the correponding lags are a lot smaller, as seen 
in Figure \ref{relccf2}. Perhaps, one should have used as an incident
light curve one with substantially more power in the high frequencies 
than the observed UV light curve, as done in Rokaki \& Magnan (1992). 
However, in the absence of such data one can only speculate.

\section{Discussion, Conclusions}

In the sections above we  explored the observational effects 
of X-ray reprocessing by a geometrically thin, optically thick 
accretion disk in AGN. We assumed in doing this that a point-like
X-ray source irradiates the accretion disk (approximated by an 
infinite plane) from a given distance above it; the disk then re emits 
the locally incident X-ray flux in black body form, much in the 
way considered for producing the BBB spectrum. Our goal has been to 
compare our results to the observations of the recent mulitwavelength 
campaign of NGC 7469, which sampled its optical, UV and X-ray light 
curves at a rate sufficiently high to allow meaningful
measurements of the lags between these components. We searched 
the relevant parameter space to examine whether it is possible to 
account for the observed correlated variability in the optical -- 
UV  bands as the result of reprocessing of the {\sl observed} X-rays 
by an accretion disk. This is a relevant question, as this process 
was suggested to be the origin of the optical -- UV lags in previous, 
less well sampled (in time and wavelength) campaigns. 

Our ``bottom line" results are summarized by Figure \ref{relccf}
which exhibits the relative lags of our model light curves between 
the UV (\l 1315 \AA) and the optical (\l 6962 \AA) bands, in order 
to compare directly with the same quantity as determined by the NGC 
7469 campaign (see figure 5 in Collier et al. 1998): The model 
lags are much too short to account for the observations under
the assumptions of our model (i.e. that they are due to X-ray 
reprocessing), for any reasonable value of the black hole mass
$M$ or the angle $\theta$. The model lags are shorter than those
observed by, at least, a factor five, for any reasonable value
of the black hole mass (or, equivalently, height of the X-ray 
source above the disk). However, the lags reported by Collier
et al. (1998) could be accounted within the model, provided that
the variability is due to the reprocessing of a continuum component 
other than the X-rays, with a light curve similar to that observed
in the UV.

This outcome, as indicated also by the detailed X-ray reprocessing 
model light curves, is the result of the short distances associated
with black hole - accretion disk systems of these models. 
For reasonable values of the black hole mass, $M$, 
the \l 1315 \AA~ light curves track almost identically that of 
X-rays. Even for the optical wavelength, \l 6962 \AA, the reprocessed
light curves follow quite closely those of the X-rays as seen in
the relevant figures and as shown by the autocorrelation functions. 
We have also produced model light curves for two additional 
wavelengths, namely \l 15000 \AA~ and \l 30000 \AA, both well outside
the range of observations, in order to explore at what wavelengths
our model could yield light curves with autocorrelation functions 
similar to those observed in the UV. It was only for the longest 
wavelength (\l 30000 \AA) and for $M = 10^8 \, {\rm M}_{\odot}$
that we were able to produce autocorrelation functions resembling 
that associated with the light curve of NGC 7469 at \l 1315 \AA. 
However, in this last case
the reprocessed emission comes from such large radii that the 
resulting RMS variability is much smaller than that the 
UV observations indicate.

In view of our model light curves of Figures \ref{m7lcurv}, 
\ref{m789lcurv}, \ref{m7m8_2lcv} and also as discussed in Nandra 
et al. (1998), the correlated X-ray --  UV variability of NGC 7469 
appears really puzzling. One  might argue that the variability in 
the UV -- optical bands is intrinsic to that of 
the accretion disk itself; as argued in Collier et al. (1998) the
wavelength dependence of the UV -- optical lags are consistent
with this view. However this would present the following 
two problems: (a) the 
observed lag time scales are too short for the standard accretion
disk models  (as discussed in the introduction); (b) given the 
roughly equal luminosities in the X-ray and UV bands, there should
exist some evidence of a high frequency component in the UV light 
curve due to the (expected) reprocessing of the observed X-rays, 
should the geometry assumed in the present note be valid.
Perhaps this last constraint could be alleviated if the albedo
of the disk were high, i.e. ${\cal A} \simeq 0.9$, as this could 
make the amplitude of the reprocessed component very hard to discern. 
It has been suggested recently that this may very well be the case 
under certain conditions in X-ray illuminated disks (Nayakshin, 
Kazanas \& Kallman 1999). In this case however, the presence of such 
a reflected component with amplitude similar to that of the intrinsic
X-ray flux should be apparent in the X-ray autocorrelation function 
at lags $\tau \sim R/c$, where $R$ is the height of the X-ray
source above the disk. Allowing for additional speculation to 
remove this last conundrum, one could argue that the X-ray emission
comes from magnetic loops (Galeev, Rosner, Vaiana 1979) of sizes
smaller than the X-ray sampling time multiplied by the speed of 
light (i.e. $R \ll 10^{14}$ cm). Such a solution appears contrived
and it still does not explain the origin of the UV -- optical lags.

In conclusion, the observations of the correlated multiwavelength 
variability of NGC 7469 seems  to be grossly incompatible with the 
simplified model examined herein, namely that this variability is
due to reprocessing of the observed X-rays by an optically thick,
geometrically thin disk, with the X-ray source approximated
by a  point source at a given  distance above this disk. It is not
apparent to us how simple modifications to this model could 
lead to results compatible with these observations. The connection 
between the X-ray and the longer wavelength (BBB) components appears to
be much less direct than allowed through this very simplified model,
which suggests much more rapid variability, yet these two spectral 
components should be somehow related, given their similar overall 
variability amplitudes.

It is not known whether the correlations of the multiwavelength 
variations observed in NGC 7469 are a general AGN property 
or particular to this specific object. These intriguing results 
suggest that additional studies with similar wavelength coverage and 
sampling rates are badly needed in order to establish whether
our present, general notions of AGN structure are indeed sound 
or in need of a major revision. 
Our models and their comparison to observations should be 
viewed only as a case in support of the argument made in the 
introduction, namely that spectral fits alone are unable to provide 
unequivocal information concerning the structure of AGN and generally
of sources powered by accretion onto compact objects. Time variability 
observations and successful modeling are an absolutely necessary 
supplement to the spectral studies. 

We would like to thank P. Nandra for providing us with the X-ray 
light curve of NGC 7469. We would also like to thank Sergei Nayakshin,
Hagai Netzer and Brad  Peterson for several interesting discussions.

\end{document}